\documentclass[twocolumn,showpacs,preprintnumbers,amsmath,amssymb]{revtex4}
\usepackage{graphicx}
\usepackage{dcolumn}

\def\be{\begin{equation}}
\def\ee{\end{equation}}

\begin{document}

\title{Exact Solvability of the two-photon Rabi Hamiltonian}

\author{I.Trav\v enec$^*$\\
Institute of Physics, Slovak Academy of Sciences, D\'ubravsk\'a cesta 9,
842 28 Bratislava, Slovakia}

%\maketitle

\begin{abstract} 

Exact spectrum of the two-photon Rabi Hamiltonian is found, proceeding in full analogy
with the solution of standard (one-photon) Rabi Hamiltonian, published by Braak 
in Phys. Rev. Lett. 107, 100401 (2011).
The Hamiltonian is rewritten as a set of two differential equations.
Symmetries that get hidden after further treatment are found. 
One can plainly see, how the Hilbert space splits into four disjunct subspaces, categorized 
by four values of the symmetry parameter $c=\pm1,\pm i$. 
There were only two values $\pm1$ for the standard Rabi model.
Four analytic functions are introduced by a recurrence scheme for the coefficients of their
series expansion.
All their roots yield the complete spectrum of the Hamiltonian.
Eigenstates in Bargmann space are also at disposal.

\end{abstract}
\pacs{03.65.Ge, 02.30.Ik, 42.50.Pq}

\maketitle

\bigskip

\section{Introduction}
There were many trials to solve the model called (in quantum optics) Rabi Hamiltonian exactly,
until Braak \cite{Braak} recently succeeded to do so. 
The same model is known under several pseudonyms, e. g. single-mode spin-boson system, 
Jaynes-Cummings model without rotating wave approximation and others; a brief survey 
can be found in \cite{ASB}.
Let us introduce a more general form of Rabi Hamiltonians \cite{Rabi}, describing the interaction 
between a bosonic mode with energy $\omega$ and a two-level atom with level spacing $\omega_0$
\begin{equation} \label{e1}
\hat{H}^{(m)} = \frac{\omega_0}{2}\sigma^z+\omega b^\dagger b
+g(\sigma^+ +\sigma^-)\left[(b^\dagger)^m+b^m\right],
\end{equation}
where $m=1,2,\ldots$; $g$ is the interaction constant, $\sigma^z$, $\sigma^{\pm}$ are 
the Pauli matrices, $b^\dagger$ and $b$ are boson creation and annihilation operators, 
respectively. 
For the most studied $m=1$ case Braak \cite{Braak} recently presented an exact algebraic
solution; we are going to solve the $m=2$ model in full analogy.

Before Braak's general solution, some special points, called also the Juddian ones,
were known to be exactly solvable. 
This means that for some constraints on the Hamiltonian parameters, one can find both 
an eigenfunction and its eigenenergy. 
It doesn't give the complete spectrum, just one excited eigenstate. 
This was shown at first for the standard $m=1$ Rabi Hamiltonian \cite{Judd}.
Later the same picture, of course with different constraints and eigenvalues was shown for $m=2$, 
i. e. the two-photon Rabi Hamiltonian \cite{E-B}.
These special points manifest themselves as cross-sections of general solutions and they will
serve as a check of our results.

It is also known that the parameters of the two-photon Rabi model are restricted by 
$|4g|<\omega$, otherwise the eigenfunctions are not normalizable \cite{E-B,Dolya}.

\section{Exact algebraic solution}

The first step of the solution is going over to Bargmann space \cite{Barg}, introducing complex
variable $z$, where the bosonic operators simplify to

\be\label{e2}
b\to \frac{\partial}{\partial z},\qquad b^\dagger\to z.
\ee

The eigenfunctions have to be analytic in the whole complex plane.
Let us suppose that the solution of the stationary Schr\"odinger equation is a vector composed 
of two analytic functions $(\psi_1(z),\psi_2(z))^T$. 
We can insert eq.(\ref{e2}) into the Hamiltonian (\ref{e1}), case $m=2$, and using the well-known
$2 \times 2$ Pauli matrices, we obtain a coupled system of second-order differential equations

\begin{eqnarray} \label{e3}
& 2g\psi_2''+\omega z \psi_1'+2gz^2\psi_2+(\frac{\omega_0}{2}-E)\psi_1 = 0  \nonumber \\ 
& 2g\psi_1''+\omega z \psi_2'+2gz^2\psi_1-(\frac{\omega_0}{2}+E)\psi_2 = 0,
\end{eqnarray}
where $'$ means the derivative with respect to $z$ and $E$ is the energy from the Hamiltonian's spectrum.
For better symmetrization it is useful to go over to a linear combination of the considered
functions, namely $\phi_1(z)=\psi_1(z)+\psi_2(z)$ and $\phi_2(z)=\psi_1(z)-\psi_2(z)$.
Thus we get the set

\begin{eqnarray} \label{e4}
& 2g\phi_1''+\omega z \phi_1'+(2gz^2-E)\phi_1+\frac{\omega_0}{2}\phi_2 = 0  \nonumber \\ 
& -2g\phi_2''+\omega z \phi_2'-(2gz^2+E)\phi_2+\frac{\omega_0}{2}\phi_1 = 0. 
\end{eqnarray}

The next step is to find important symmetries present in the set (\ref{e4}) before they become less 
clear after some transformation followed by a series expansion. 
Here we prefer a somewhat different way than Braak \cite{Braak}, but a very obvious one.
We will perform two transformations of the variable $z$.
The first one is $z\to -z$. 
One can easily see, that it leaves the set (\ref{e4}) unchanged.
Simple inspection shows, that this implies common parity of both $\phi_1$ and $\phi_2$.
Plainly speaking, they must be either both even

\begin{eqnarray} \label{e5}
& \phi_1(-z)=\phi_1(z)  \nonumber \\ 
& \phi_2(-z)=\phi_2(z),
\end{eqnarray}
or alternatively, simultaneously odd

\begin{eqnarray} \label{e6}
& \phi_1(-z)=-\phi_1(z)  \nonumber \\ 
& \phi_2(-z)=-\phi_2(z).
\end{eqnarray}

The second applied transformation is $y=i z$. The set (\ref{e4}) becomes

\begin{eqnarray} \label{e7}
-2g\frac{{\rm d}^2 \phi_1}{{\rm d}y^2}+\omega y \frac{{\rm d} \phi_1}{{\rm d}y}-(2g y^2+E)\phi_1+\frac{\omega_0}{2}\phi_2 = 0 & \nonumber \\ 
2g\frac{{\rm d}^2 \phi_2}{{\rm d}y^2}+\omega y \frac{{\rm d} \phi_2}{{\rm d}y}+(2g y^2-E)\phi_2+\frac{\omega_0}{2}\phi_1 = 0. &
\end{eqnarray}

Now the functions $\phi_1$ and $\phi_2$ evidently swapped their places, thus up to some 
common multiplicative constant:

\begin{eqnarray} \label{e8}
& \phi_1(i z)= c\ \phi_2(z)  \nonumber \\ 
& \phi_2(i z)= c\ \phi_1(z).
\end{eqnarray}

Possible values of the parameter $c$ can be found by inserting $i z$ instead of $z$ and we get
$\phi_1(-z)=c\ \phi_2(i z) = c^2\ \phi_1(z)$ and analogously $\phi_2(-z)= c^2\ \phi_2(z)$.
Hence $c^2=1$ from eq.(\ref{e5}) or $c^2=-1$ from eq.(\ref{e6}). The symmetry parameter can
acquire four values, $c=\pm 1,\pm i$.
In the standard $m=1$ Rabi model such parameter had only two values $\pm1$ and the transformation
$z\to-z$ was sufficient. 

Finally let us introduce four functions $G_c(z,E)$ whose roots with respect to $E$ will be used to fulfill
eq.(\ref{e8}):

\begin{eqnarray} \label{e9}
& G_+(z) = \phi_2(i z)-\phi_1(z)  \nonumber \\ 
& G_-(z) = \phi_2(i z)+ \phi_1(z)  \nonumber \\ 
& G_i(z) = i \phi_2(i z)+ \phi_1(z)  \nonumber \\ 
& G_{-i}(z) = i\phi_2(i z)- \phi_1(z).
\end{eqnarray}
In the last two cases the lower of eqs.(\ref{e8}) was multiplied by $i$ in order to keep the $G_c$ functions 
real. 
They share the common parity of their $\phi_{1,2}$ functions. 
The upper of eqs.(\ref{e8}) is then no more independent.
The complete discrete spectra will be given by all roots of all $G_c(z,E)$ functions, again in full
analogy with the standard Rabi model \cite{Braak}.
Of course, the roots are meant with respect to $E$ and they are independent of any chosen $z$.

Having categorized the symmetries, we return to the set (\ref{e4}) and perform the transformation
$\phi_{1,2}= e^{-\kappa z^2}\bar\psi_{1,2}$. We get

\begin{eqnarray} \label{e10}
& 2g\bar\psi_1''+(\omega -8 g \kappa) z \bar\psi_1'-(4g\kappa+E)\bar\psi_1+
\frac{\omega_0}{2}\bar\psi_2 = 0  \nonumber \\ 
& -2g\bar\psi_2''+(\omega+8g\kappa)z\bar\psi_2'-(4\omega\kappa z^2 - 4g\kappa+E)\bar\psi_2 + \nonumber \\
& + \frac{\omega_0}{2}\bar\psi_1 = 0, 
\end{eqnarray}
where we used the parameter $\kappa$ to simplify the first equation by removing the term
$(8g\kappa^2-2\omega\kappa +2g)z^2\bar\psi_1$, thus specifying its value

\be\label{e11}
\kappa = \frac{\omega - \sqrt{\omega^2-16 g^2}}{8g}.
\ee

In fact there should be $\pm$ in front of the square root, but the plus sign would make $\kappa$
divergent in the limit $g\to 0$, which is physically not reasonable. 
Notice that $\kappa$ remains real under the above mentioned restriction $4 |g| < \omega$.
There is also a further analogy with the standard Rabi model, where the special case 
$\omega_0=0$ is exactly solved with help of the coherent state 
$\exp(\pm 2g/\omega\ b^\dagger)|0\rangle$, here $|0\rangle$ is the lowest bosonic state
\cite{Bish,T-S}. 
In our notation, Braak performs the transformation $\phi_{1,2}\propto\exp(-2 g z/\omega)\bar\psi_{1,2}$, recall $b^\dagger\to z$ in (\ref{e2}). 
The $m=2$ Rabi Hamiltonian solution at $\omega_0=0$ involves the squeezed vacuum \cite{Brif} 
term $\exp(\pm \kappa\ {b^\dagger}^2)|0\rangle$.
We will return to this case later.

In the next step we expand the functions $\bar\psi_{1,2}$:

\begin{eqnarray} \label{e12}
\bar\psi_1(z)= \sum_{n=-\infty}^\infty Q_n(E)z^n  \nonumber \\ 
\bar\psi_2(z)= \sum_{n=-\infty}^\infty K_n(E)z^n.
\end{eqnarray}

To keep the solutions analytic we expect $Q_n(E)=K_n(E)=0$ for $n<0$ \cite{Braak}. 
Inserting these expressions into eqs.(\ref{e10}) we get the iteration scheme

\begin{eqnarray} \label{e13}
2g(n+2)(n+1)Q_{n+2} + \left[(\omega-8g\kappa)n-4g\kappa-E\right]Q_n
\nonumber\\ + \frac{\omega_0}{2}K_n= 0  \nonumber \\ 
-2g(n+2)(n+1)K_{n+2} + [(\omega+8g\kappa)n+4g\kappa-E]K_n
\nonumber\\ - 4\omega\kappa K_{n-2}+\frac{\omega_0}{2}Q_n = 0.\ \ \ \ 
\end{eqnarray}

Notice that the indexes differ by 0, 2 or 4, thus only the coefficients $Q_n,$ $K_n$ with
common parity will be non-zero, either for even $n=0,2,4\ldots$, eq.(\ref{e5}), or odd 
$n=1,3,5\ldots$, eq.(\ref{e6}). 
The prefactor $ e^{-\kappa z^2}$ doesn't spoil the parity.
The starting point of our iteration scheme is either at $n=0$ or $n=1$.
Let us first have a look on the case $n=0$ and the symmetry parameter $c=1$.
Upper eq.(\ref{e8}) $\phi_1(i z)= \phi_2(z)$ at $z=0$ implies $Q_0=K_0$, which can itself be a function
of Hamiltonian parameters, serving as normalization constant for the eigenfunctions.
But as we are interested only in the roots of $G_+(z,E)$, this constant becomes an unimportant
multiplication factor and we can choose

\be\label{e14}
Q_0=1,\qquad K_0=1.
\ee

The second case is $c=-1$. We have $\phi_1(i z)= -\phi_2(z)$ and again at $z=0$

\be\label{e15}
Q_0=1,\qquad K_0=-1.
\ee

The third case with $c=i$ is only a bit more complicated. The symmetry condition from eq.(\ref{e8}) 
is now $\phi_1(i z)= i \phi_2(z)$ and its value at $z=0$ is rather trivial $Q_0=K_0=0$. 
But we will compare the first derivatives w. r. t. $z$, i. e. $\phi_1'(i z)= i \phi_2'(z)$ at $z=0$
and we get $i Q_1=i K_1$, thus now our starting point is

\be\label{e16}
Q_1=1,\qquad K_1=1.
\ee

At last for $c=-i$ we get

\be\label{e17}
Q_1=1,\qquad K_1=-1.
\ee

Concluding this part we can see that the Hilbert space of eigenfunctions splits into four disjunct
subspaces. 
The corresponding eigenvalues can be found separately as roots of four $G_c(E)$ functions, 
eq.(\ref{e9}).
We substitute $\phi_{1,2}=e^{-\kappa z^2}\bar\psi_{1,2}$, the coefficients of expanded $\bar\psi_{1,2}$ 
functions are found applying the iteration scheme (\ref{e13}) subsequently with four starting points, eqs.(\ref{e14}-\ref{e17}).
The coefficients not defined by eq.(\ref{e13}) are zero because of parity demands.
$K_{-2}=K_{-1}=0$ as well.

Before proceeding to numerical calculations, we ought to mention several special cases, where the
exact solution was already known.
They will serve as a check of our general solution.

\section{Some exactly known cases}

\subsection{Case $g=0$}

If the interaction constant is zero, the system separates into independent two-state atom
with energy levels $\pm\omega_0/2$ and a phonon with the modes $N \omega$, where $N=0,1,2,\ldots$
Thus the overall energy is $\pm\omega_0/2+N \omega$. 
It is instructive to see how these values split into four groups as roots of four $G_c$ functions.
Therefore we will solve this simple case explicitly.
We return to the original functions $\psi_{1,2}$, as the set of equations (\ref{e3}) decouples
for $g=0$ and the two independent solutions are found easily

\begin{eqnarray} \label{e18}
\psi_1=C_1 z^{\frac{2E-\omega_0}{2\omega}}=C_1 z^{k'} \nonumber \\ 
\psi_2=C_2 z^{\frac{2E+\omega_0}{2\omega}}=C_2 z^k,
\end{eqnarray}
where we denoted the exponents by $k'$ and $k$, as the parity conditions (\ref{e5})
and (\ref{e6}) are common for $\phi_{1,2}$ and $\psi_{1,2}$ functions.
They force $k'$ and $k$ to be integer and the analyticity demands make them non-negative.
The energies $E=-\omega_0/2+k \omega$ and $E=\omega_0/2+k' \omega$ should be common, but they
are generally different, thus the overall solutions will have either $C_1=0$ or $C_2=0$.

Let us first analyze the case $(0,\psi_2)^T$ with the energy $E=-\omega_0/2+k \omega$.
The linear combinations $\phi_1=\psi_1+\psi_2=C_2 z^k$ and $\phi_2=\psi_1-\psi_2=-C_2 z^k$.
The solutions for the functions $G_c=0$ finally yield

\begin{eqnarray} \label{e19}
G_-=-C_2 (i z)^k+C_2 z^k=0\Rightarrow \  k=0,4,8,\ldots \nonumber \\ 
G_+ =-C_2 (i z)^k-C_2 z^k=0\Rightarrow \ k=2,6,10,\ldots \nonumber \\ 
G_{-i} =-i C_2 (i z)^k+C_2 z^k=0\Rightarrow \ k=3,7,11,\ldots \nonumber \\ 
G_i =-i C_2 (i z)^k-C_2 z^k=0\Rightarrow \ k=1,5,9,\ldots 
\end{eqnarray}

The second solution $(\psi_1,0)^T$ with $\phi_1=\phi_2=C_1 z^{k'}$ and eigenenergies 
$E=\omega_0/2+k' \omega$ gives

\begin{eqnarray} \label{e20}
G_-=C_1 (i z)^{k'}+C_1 z^{k'}=0\Rightarrow \ k'=2,6,10,\ldots \nonumber \\ 
G_+=C_1 (i z)^{k'}-C_1 z^{k'}=0\Rightarrow \ k'=0,4,8,\ldots \nonumber \\ 
G_{-i}=i C_1 (i z)^{k'}+C_1 z^{k'}=0\Rightarrow \ k'=1,5,9,\ldots \nonumber \\ 
G_i=i C_1 (i z)^{k'}-C_1 z^{k'}=0\Rightarrow \ k'=3,7,11,\ldots 
\end{eqnarray}

We will later analyze mainly the cases when $\omega_0$ and $\omega$ are comparable
and the eigenenergies as roots of $G_c$ functions reorganize as follows:

\begin{eqnarray} \label{e21}
& G_-: & \quad -\frac{\omega_0}{2},\quad \frac{\omega_0}{2}+2\omega,
\quad -\frac{\omega_0}{2}+4\omega,\ldots \nonumber\\
& G_+: & \quad \frac{\omega_0}{2},\quad -\frac{\omega_0}{2}+2\omega,
\quad \frac{\omega_0}{2}+4\omega,\ldots \nonumber\\
& G_{-i}: & \quad -\frac{\omega_0}{2}+\omega,\quad \frac{\omega_0}{2}+3\omega,
\quad -\frac{\omega_0}{2}+5\omega,\ldots \nonumber\\
& G_i: & \quad \frac{\omega_0}{2}+\omega,\quad -\frac{\omega_0}{2}+3\omega,
\quad \frac{\omega_0}{2}+5\omega,\ldots 
\end{eqnarray}
We can see that real values of the symmetry parameter $c$ are connected with even number of phonon
excitations $N$ whereas the imaginary $c$ is coupled to odd $N$. 
The global ground state is always given by the lowest root of $G_-$, which is true also 
for non-zero $g$.

\subsection{Case $\omega_0 = 0$}

If the gap $\omega_0$ between atomic levels disappears, all eigenenergies become twice degenerate.
They are exactly known and now the complete spectrum is given by \cite{E-B}

\be\label{e22}
 E=-\frac{\omega}{2}+(n+\frac{1}{2})\Omega\omega \qquad      n=0,1,2,\ldots
\ee
where another dimensionless quantity was introduced
\be{\label{e23}}
\Omega=\sqrt{1-\frac{16 g^2}{\omega^2}}=1-\frac{8g\kappa}{\omega}.
\ee

Let us reproduce this result. 
We return to the $\phi_{1,2}$ functions, because now the system of equations (\ref{e4})
decouples for $\omega_0=0$ and the general solutions are

\begin{eqnarray} \label{e24}
&\phi_1=\exp\left[{(\kappa-\frac{\omega}{4 g})z^2}\right]. \nonumber\\
&\left[C_{11}H_{-n-1}\left(\sqrt{\frac{\omega\Omega}{4 g}}z\right)+ 
C_{21}\ {}_1F_1\left(\frac{n+1}{2},\frac{1}{2},\frac{\omega\Omega}{4 g}z^2\right)\right] \nonumber \\
&\phi_2=\exp\left({\kappa z^2}\right). \nonumber\\
&\left[C_{12}H_n\left(\sqrt{\frac{\omega\Omega}{4 g}}z\right)+ 
C_{22}\ {}_1F_1\left(-\frac{n}{2},\frac{1}{2},\frac{\omega\Omega}{4 g}z^2\right)\right],
\end{eqnarray}
exploiting the well known Hermitian polynomials $H_n$ and hypergeometric function ${}_1F_1$. 
Further we introduced the quantity $n=(\omega+2E)/(2\Omega\omega)-1/2$.
It is nothing else but eq.(\ref{e22}) reversed.
Thus if the parity demands force $n$ to be non-negative integer again, the spectrum is reproduced.
Let us show it in detail at least for the simpler case of $n$ even.
We make use of a formula relating Hermitian polynomials and hypergeometric function,
valid for $n=0,2,4,\ldots$

\be{\label{e25}}
\frac{H_n(i q z)}{2^{n/2}(n-1)!!}=i^n e^{-q^2 z^2} {}_1F_1\left(\frac{n+1}{2},
\frac{1}{2},q^2 z^2\right), 
\ee
whereas for any $n$, including non-integer one, the Kummer transformation \cite{P-Z} gives

\be\label{e26}
{}_1F_1\left(-\frac{n}{2},\frac{1}{2},-y\right)=e^{-y} {}_1F_1\left(\frac{n+1}{2},\frac{1}{2},y\right). 
\ee

First of all we set $C_{11}=0$, because the Hermitian polynomial with negative index has no parity,
whereas the rest of the solutions (\ref{e24}) are even functions for $n$ even.
We define $q^2=\omega\Omega/(4 g)$, $y=q^2 z^2$ and require $G_{\pm}=0$ using eqs. (\ref{e23}) - (\ref{e26}).
We get 

\be\label{e27}
(-2)^{n/2}(n-1)!!C_{12}+C_{22}\mp C_{21} = 0.
\ee
For $n$ non-integer (or odd integer), the term with the Hermitian polynomial would become
complex, but our functions $\phi_2(i z)$ and $G_\pm$ are real - hence we set also $C_{12}=0$.
We already mentioned that one parameter can be chosen, say the integration constant $C_{21}=\phi_1(0)=1$.
Thus $C_{22}=\phi_2(0)=\pm 1$ and the two $G_c$ functions are identical, $G_+=G_-$. 
One can compare the series expansion of these $\phi_{1,2}(z)$ solutions with those given 
by the scheme (\ref{e13}), starting points (\ref{e14}) or (\ref{e15}) and see that in fact we 
managed to perform the complete sum of eqs. (\ref{e12}).

Having exact formulas for some $\phi_{1,2}$ and  $G_c(z,E)$ functions at disposal allows us to
make several observations.
In the on-line supplement to ref. \cite{Braak}, Braak reports some problems with the radius
of convergence $R$ of his series in $z$. 
$R$ seemed to be finite in some cases, though analyticity in the whole complex plane of $z$ is required. 
To keep the series analogous to our eqs.(\ref{e12}) convergent, as a necessary condition, the ratio $K_{n+1}/K_n$ had to go to zero for $n\to\infty$. 
If it was non-zero, $R$ became finite.
For the special case $\omega_0=0$ we can calculate this ratio explicitly.
The coefficients $Q_n$ with $n=2k$ become rather simple

\be\label{e28}
Q_{2k}=\frac{1}{(2 g)^k (2k)!}\left(E-\varepsilon_{2k-2}\right)\left(E-\varepsilon_{2k-4}\right)
\ldots\left(E-\varepsilon_0\right),
\ee
where $\varepsilon_n$ are eigenenergies from eq.(\ref{e22}). En route we see that the
energy $E=\varepsilon_n$ terminates $\bar\psi_1(z)$ to a (Hermite) polynomial. 
The ratio 

\be\label{e29}
Q_{2k+2}/Q_{2k}\approx -\frac{\omega\Omega}{2g}\frac{1}{2k}\to 0,\qquad k\to\infty
\ee
as required. 
The equations for $K_{2k}$ are not so simple, thus we resort to the exact solution of $\phi_2$
and find

\be\label{e30}
K_{2k+2}/K_{2k}\approx \frac{\omega\Omega}{2g}\frac{1}{2k}\to 0,\qquad k\to\infty.
\ee
For $n$ odd and especially for non-zero $\omega_0$ we performed at least numerical study of 
$K_n$, $Q_n$ from the scheme (\ref{e13}) and found that their ratios are also proportional
to $1/n$ in leading term.
Hence we experience no problems with analyticity of the expanded functions including even very
large $z$.

The next remark concerns the practical numerical calculations of the roots $G_c(z,E)=0$,
independent of $z$.
It turns out that such calculations are numerically more stable for large $z$, which is allowed 
by the previous notion.
One can exploit the large-$z$ asymptotic for the $\phi_1(z)$ contribution to $G_\pm$
given by \cite{P-Z}

\be
{}_1F_1\left(a,b,y\right)=\frac{\Gamma(b)}{\Gamma(a)}e^y y^{a-b}\left[1+{\cal O} \left(\frac{1}{y}\right)
\right] \qquad y\gg 0,
\ee
which is analytic for non-negative integer power $a-b=n/2$, i. e. $n=0,2,\ldots$ as expected. 
$\phi_2(i z)$ gives the same after Kummer transformation (\ref{e26}).

Further we return to the complete spectrum in eq.(\ref{e22}).
For $|g|\to\omega/4$ the quantity $\Omega\to 0$ and the energy becomes infinitely many times
degenerate.
This point is physically unsound, though well defined in the sense of a limit.

Concluding, the eigenvalue of eq.(\ref{e22}) for $n=0$ coincides with mutually equal lowest roots 
of $G_-$ and $G_+$, for $n=1$ it is the lowest root of both $G_{-i}$ and $G_i$, for $n=2$ the second 
lowest root of both $G_-$ and $G_+$, etc.

\subsection{Special cases with non-zero $\omega_0, \omega, g$}

Let us now recall the result of Emary and Bishop \cite{E-B}, who found a set of isolated
solutions for our model. 
We are not going to rederive it, but the basic fact is that under some constraint on Hamiltonian's
parameters and energy $E$, at least one of the original eigenfunctions $\psi_{1,2}$ becomes 
a product of some exponential function and of a polynomial.

The main statement is that there exist exactly known eigenstates with eigenenergies

\begin{equation}{\label{e31}}
E=-\frac{\omega}{2}+\left( N+\frac{1}{2}\right)\Omega\omega
\end{equation}
if the following conditions are fulfilled

\begin{eqnarray} \label{e32}
2-6\Omega^2+\frac{\omega_0^2}{4\omega^2}=0,\qquad N=2, \nonumber\\
6-10\Omega^2+\frac{\omega_0^2}{4\omega^2}=0,\qquad N=3, \nonumber\\
8(3-30\Omega^2+35\Omega^4)+2(7-17\Omega^2)\frac{\omega_0^2}{4\omega^2} \nonumber\\
+\frac{\omega_0^4}{16\omega^4}=0,\qquad N=4,
\end{eqnarray}
etc. We will proceed so that we choose the values of $\omega_0$ and $\omega$, than we gradually 
change $g$. 
Eq.(\ref{e32}) yields value(s) of $g$ and followingly the energy is got from eq.(\ref{e31}).
These solutions will manifest themselves as crossing points of appropriate roots of $G_-$ and $G_+$ 
in the case of $N$ even, or crossings of $G_{-i}$ and $G_i$ roots in the case of $N$ odd.
This is again analogous with the standard $m=1$ Rabi model, where the Juddian points appeared 
as crossings of the roots of $G_-^{(1)}$ and $G_+^{(1)}$ \cite{Braak}.
Recall that contrary to eq. (\ref{e22}), this is only one known excited state for chosen set of 
Hamiltonian parameters, not the complete spectrum.

\begin{figure} [t]
\includegraphics[scale=0.33]{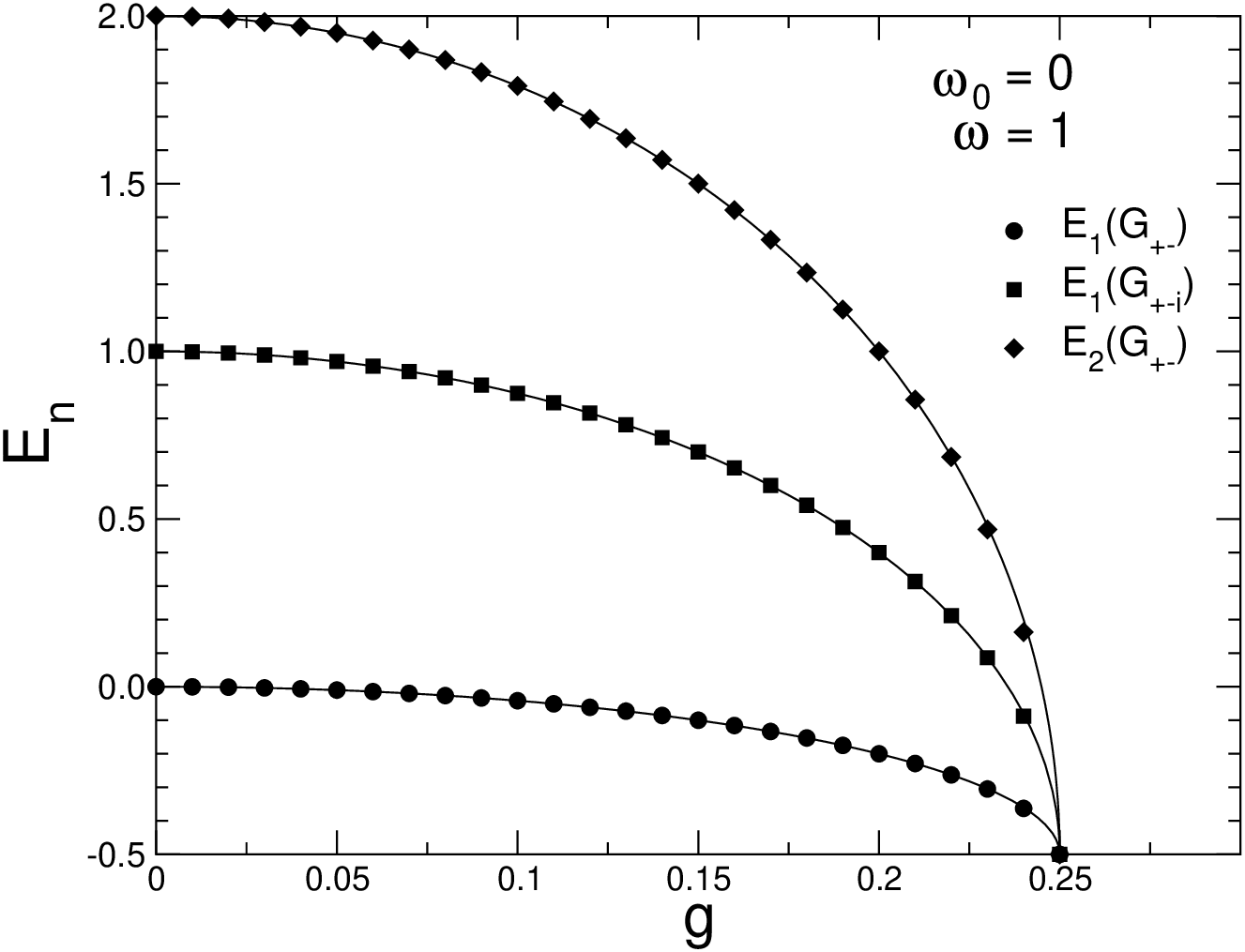}
\caption{\label{fig1}Eigenenergies as roots of $G_c(E)$ functions for $\omega_0=0$ and $\omega=1$. Full lines
are exact values.}
\end{figure}

\section{Numerical results}

The resulting energies as roots of $G_c$ functions should not depend on $z$; this would be true
after summing infinite number of summands in eq.(\ref{e12}).
In numerical calculations we truncate the sum and hope that the higher powers are not significant.
We cannot use very small $z$, because of numerical instability.
Somewhat surprisingly, we can use as large values as $z=1000$ or even $z=10^4$ without real change
of the roots. 
This fact was already noted for exactly solvable cases, nevertheless one even doesn't have to use 
compromise medium values despite of the truncated expansions.
Using the symbolic program Mathematica, we could sum up to $z^L$ term with $L=34$ for even
solutions or $L=35$ for the odd ones.
If the values of a root with smaller $L$ converged to the same value, we accept it.
It turns out that for smaller values of $|g|$ the convergence is excellent, but it becomes poorer
as we approach the maximal possible value, i. e. if $|g|\to \omega/4$.
The interval of $g$ in our plots is limited to $0\le g \le w/4$.
It is known that although the eigenfunctions differ after changing the sign of $g$, the eigenenergies
remain the same, i. e. there is a mirror symmetry $E(-g)=E(g)$.

Let us test our calculations at first on the exactly solved case $\omega_0=0$.
We choose $\omega=1$.
Parts of parabolas with common top form the exact spectrum from eq. (\ref{e22}), see full lines
in Fig. 1.
It is clear, that in the vicinity of the infinitely many times degenerate point with $g=\omega/4$,
the roots of $G_c(E)$ functions become very dense and the functions themselves are quickly oscillating.
That is the reason, why even for the orders as large as $L=34$ the values of appropriate roots
did not converge completely and we have to resort to some fitting procedure, yielding better
guess of the saturation value for $L\to\infty$.
The roots are denoted so that the lowest one is $E_1(G_c)$, the second lowest one $E_2(G_c)$, etc.
We can see that the calculated eigenenergies fit the exact values almost perfectly, except for some
deviation at $g$ close to $\omega/4$ and for the higher root, in this case $E_2(G_\pm)$.

\begin{figure} [t]
\begin{center}
\includegraphics[scale=0.33]{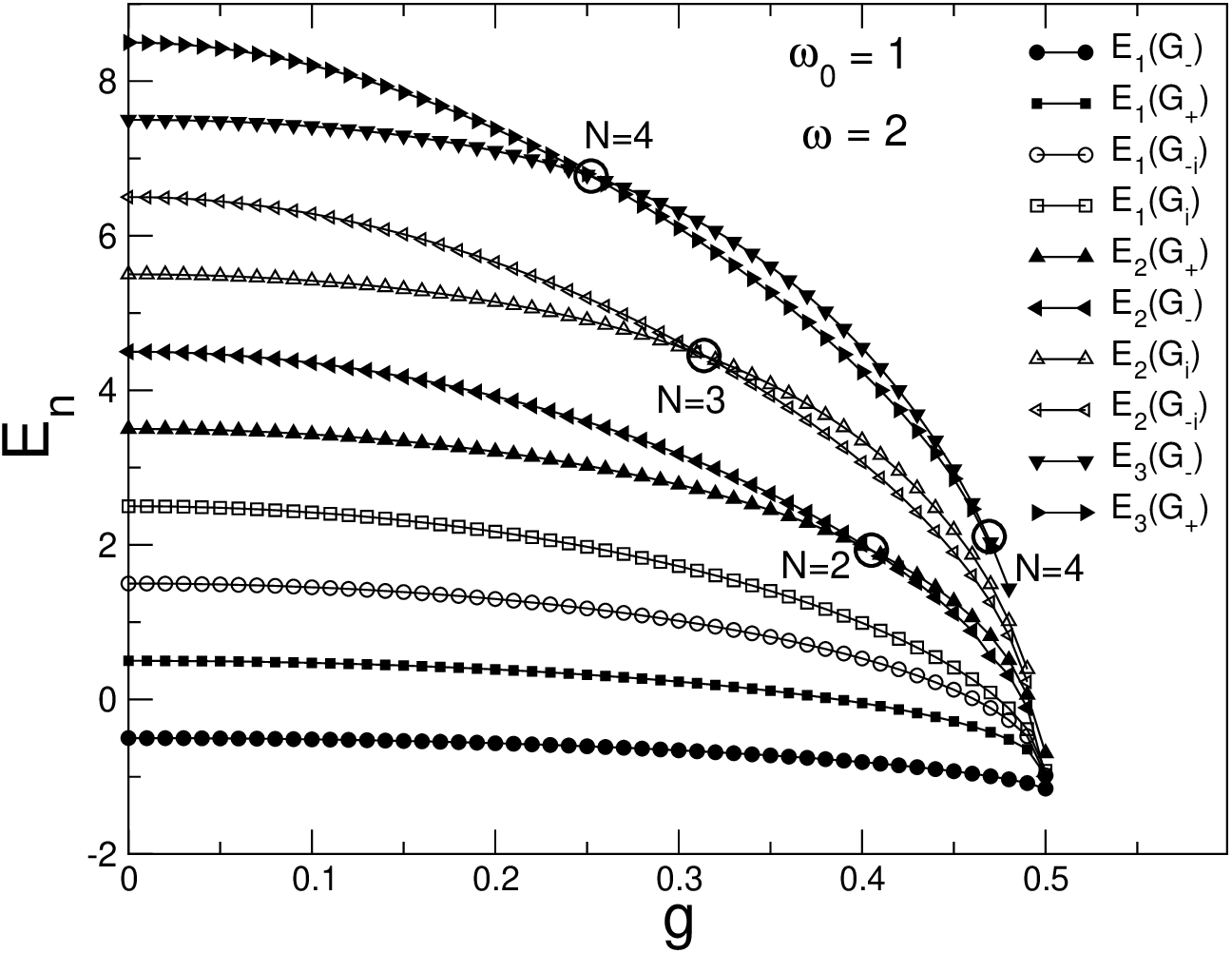}
\end{center}
\caption{Eigenenergies as roots of $G_c(E)$ functions for $\omega_0=1$ and $\omega=2$.}
\label{fig3}
\end{figure}

In Fig. 2 we present three lowest roots of $G_{\pm}$, full symbols, and two lowest roots of 
$G_{\pm i}$, smaller empty symbols.
The values $\omega_0=1$ and $\omega=2$ are chosen so that the spectrum at $g=0$ becomes
equidistant, see eq.(\ref{e21}).
The large empty circles are exact solutions of eqs.(\ref{e32}) and (\ref{e31}).
The value of $N$ is written nearby.
We can see almost perfect match with the crossings of appropriate lines.
The bottom four lines with $N=0$ and $N=1$ do not cross; lines with $N=4$ cross twice.

Fig. 3 shows the same roots denoted by the same symbols as Fig.2, just for $\omega_0=2$ and $\omega=1$.
Besides the well fitted exact crossing points, there are other crossings of lines with
different $N$, which are not exactly known.
A similar figure with in fact the same Hamiltonian parameters was already published \cite{ASB},
where the authors plot also numerical results from larger matrices diagonalization.

There is a couple of simpler analytic results, that can be got from our approach. 
So we can find e. g. the small-$g$ expansion of the (global) ground state energy $E_0=E_1(G_-)$:

\be{\label{e33}}
E_0\approx -\frac{\omega_0}{2}-\frac{8 g^2}{2\omega+\omega_0}+{\cal O}(g^4), \qquad 
g\ll \omega,\omega_0
\ee
which can be compared with similar result for the $m=1$ Rabi model: 
$E_0^{(1)}\approx-\omega_0/2-4g^2/(\omega+\omega_0)+\ldots$ \cite{T-S}.

\begin{figure} [t]
\begin{center}
\includegraphics[scale=0.33]{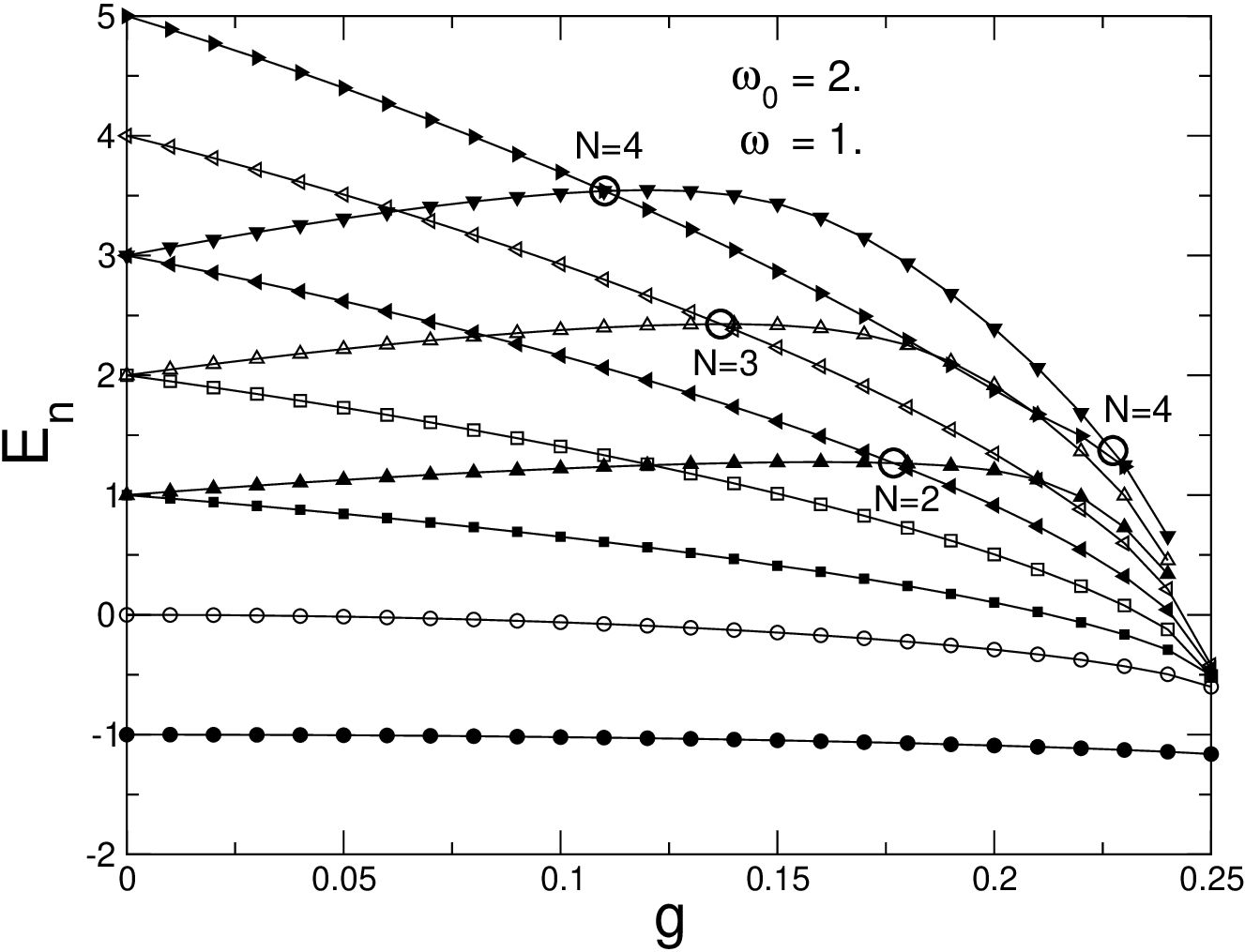}
\end{center}
\caption{Eigenenergies as roots of $G_c(E)$ functions for $\omega_0=2$ and $\omega=1$. The description 
of symbols is the same as in Figure 2.}
\label{fig2}
\end{figure}

\section{Summary}

We have found the complete spectrum of the two-photon Rabi Hamiltonian as roots of four
analytic functions $G_\pm$ and $G_{\pm i}$, in the whole parametric space.
These functions are given by the recurrence scheme (\ref{e13}) with four starting points
(\ref{e14}) - (\ref{e17}).
The unnormalized eigenfunctions in Bargmann space can be found as well, using 
$\psi_1(z)=[\phi_1(z)+\phi_2(z)]/2$ and $\psi_2(z)=[\phi_1(z)-\phi_2(z)]/2$.

In his ''Viewpoint: The dialogue between quantum light and matter'' \cite{Sol}, E. Solano states
that Braak \cite{Braak} managed to enlarge the class of exactly solvable models and
that he added the standard Rabi model on the short list of exactly solvable quantum systems.
We believe that this paper adds also the two-photon Rabi Hamiltonian on the same list.
This list can almost surely be extended further by using Braak's approach on other related models 
and yet another task for future is deeper understanding of the criteria of its applicability.

\bigskip
\acknowledgments
The author is indebted L. \v Samaj for valuable comments. This work was supported by the Grants VEGA No. 2/0049/12.

\end{document}